\documentclass[a4paper,11pt]{article}
\usepackage{jinstpub} % for details on the use of the package, please see the JINST-
\usepackage{lineno,url}

\usepackage{amsmath, tikz}
\title{Evaluating plastic scintillator performance as a substitute of LYSO in SiPM based animal PET scanners: A GEANT4 simulation analysis}

\author{Davinder Siwal$^1$, P. K. Mohanty$^2$, D. Bose$^1$, A. Jain$^2$}
\affiliation{$^1$Department of Physics, Central University of Kashmir, Main Campus, Tulmulla, Ganderbal- 191131, Jammu and Kashmir, India}
\affiliation{$^2$Tata Institute of Fundamental Research, Homi Bhabha Road, Mumbai 400005, India}
\emailAdd{dsiwal@cukashmir.ac.in}
\abstract{A systematic study is conducted to understand the coincident resolving time (CRT) for a pair of Lutetium–yttrium oxyorthosilicate (LYSO) and the plastic scintillation detector bars under the Geant4 framework. Crystals are coupled to a silicon photomultiplier single pad wafer with an appropriate optical coupling for signal readout. Pad reads the light photons undergoing optoelectronic conversion at the wafer site and generates electrical pulses with a bi-exponential shape. These signals are used to determine the trigger time stamp of back-to-back $\gamma$-rays emitted from a point source, enabling the evaluation of CRT performance at different plastic scintillator lengths. For the LYSO detector, the simulation yields the CRT response of 174 ps,  which aligns to the literature reported value for the dimensions 2 mm $\times$ 2 mm $\times$ 10 mm. To identify the plastic scintillator dimensions with an integrated $\gamma$-ray detection efficiency comparable to LYSO's photopeak efficiency at  511 keV gamma photons, various bar lengths of commercial plastic BC404 and TIFR Ooty's in-house developed plastic material are attempted in Geant4. Consequently, for both the plastic scintillators, the equivalent length (for the same cross-sectional area) was found to be  4 times that of LYSO crystal length at a threshold of 25 keV. CRT value determined for this dimension is found to be $\approx$ 300 ps for both the plastic medium. It suggests that, if an animal preclinical PET scanner is developed with plastic bars, the minimum achievable image resolution (FWHM) of $\approx$ 4.5 cm can be expected for a pair of detection elements.}

\keywords{Silicon photomultiplier, Geant4, optical photon, detection efficiency, LYSO, plastic scintillators}

\begin{document}

\maketitle
\flushbottom

\section{Introduction}
Time-of-Flight Positron Emission Tomography (TOF-PET) serves as a crucial imaging modality for precisely localizing carcinogenic regions by employing the ring configuration of fast timing and high sensitivity scintillation detectors. This technique utilizes fluorodeoxyglucose ($^{18}$F) labelled radioisotopes that accumulated at the target sites and undergo $\beta^{+}$ decay, producing positrons. These positrons rapidly annihilate with surrounding electrons, emitting back-to-back 511 keV $\gamma$-ray photons. Coincidence detection of these photon pairs by opposing detector elements establishes lines of response (LORs) that geometrically constrain the emission location. A TOF-PET system acquires numerous such LORs from multiple annihilation points within the region of interest, forming the fundamental dataset for image reconstruction. Conventional image reconstruction algorithms \cite{imagealgo1, imagealgo2} implemented on LORs can reproduces the metabolic activity mapping of the target that can be substantially enhanced by incorporating TOF information \cite{Conti2019}. Under a constrained time window up to 10 ns  \cite{zaidiPET} the scattered events can be further eliminated and it thus improves the image resolution. Traditionally, animal PET scanner employs Lutetium–yttrium oxyorthosilicate (LYSO) crystal due to their superior light output, fast decay time, good energy resolution and compatibility with silicon photomultiplier (SiPM) to make a compact array. The timing resolution of crystal bars in LYSO arrays critically influences the overall time broadening, typically 1-4 ns, observed in PET modules \cite{Zerra2017, Chen2025}. For example, studies report a coincidence time resolution (CTR) of 190 ps for 3 $\times$ 3 $\times$ 10 mm$^{3}$ LYSO crystals coupled to SensL SiPMs, which degrades to 230 ps when using with larger-area SiPMs \cite{Li2025}. Other studies for animal PET systems, measures the optimal CTR values for 4 mm$^{2}$ cross-section bars as 171.6 ps and 179 ps for 15 mm and 20 mm lengths, respectively \cite{Liu2015}. Accurate timing response of imaging crystal bar can be theoretically modeled through optical photon dispersion in scintillators, as described by Seifert \textit{et al.} \cite{Stefan2012}, whose methodology establishes the lower bound of timing resolution. We have adapted this approach to analyse timing resolution contributions from various lengths of plastic scintillators (4 mm$^{2}$ cross-section). Although plastic scintillators have not traditionally been used in commercial PET scanners, however, recent work by the Jagiellonian university group has renewed interest in their potential application in the whole-body PET scanners \cite{Baran2025}. This motivate us to evaluates the commercial BC404 and TIFR Ooty’s developed plastic scintillators \cite{ootyscint} (from cosmic ray laboratory) timing response as a potential candidates for imaging capabilities. Our findings in the article aims to facilitate the development of cost-effective animal PET systems relevant to the Indian research context. This paper is formulated as : the light transport theory is described in section \ref{sec:light}. Implementation of crystal geometry with SiPM is described in section \ref{sec:geant4_implement}. SiPM signal development is discussed in Section \ref{sec:sipm_sig}. Analysis with plastic scintillators is mentioned in Section \ref{sec:plastic_analysis}, followed by discussions in Section \ref{sec:discussion}.
\label{sec:intro}
\section{Light transport theory in scintillators}
\label{sec:light}
Detectors used in Time of Flight PET imaging utilizes array of crystal spectators arranged in a circular geometry to detect the annihilation photons of energy 511 keV. They are being emitted back-to-back from the positron emitter source having the annihilation time t$_{annih}$, that can be reasonably approximated to Gaussian distribution. Their interaction with the crystal element resulted into the emission of scintillation photons, where the time distribution is governed by the characteristic time constant of scintillation temporal profile. Let the ordered set received on the SiPM pad arranged in increasing order be $\{t_{1,1}, t_{2,1},\cdots t_{j,1}, t_{1,2}, t_{2,2} \cdots, t_{j,2}, \cdots, t_{j,k}\}$, where the first and second index represent photon and event number, respectively. Assuming each photon time stamp is a result of a Gaussian probability distribution, the observed probability function at the detection end for a given annihilation time t$_{annih}$ is given by the multivariate Gaussian density function \cite{RVinke2014} as:
\begin{equation}
P[\mathbf{t} \mid t_{\text{annih}}] = \frac{1}{\sqrt{(2\pi)^j \lvert \mathbf{\Sigma} \rvert}} \cdot \exp\left( -\frac{1}{2} (\mathbf{t} - \boldsymbol{\mu} - t_{\text{annih}})^T \mathbf{\Sigma}^{-1} (\mathbf{t} - \boldsymbol{\mu} - t_{\text{annih}}) \right)
\label{eq:probdensity}
\end{equation}
Here, $\boldsymbol{\mu}, \boldsymbol{t}$ are the vectors containing the mean arrival time and time stamp of all the indexed photons, respectively. Symbols, $\boldsymbol{\Sigma}$, $|\boldsymbol{\Sigma}|$ and $\boldsymbol{T}$ represents the covariance matrix, determinant of matrix, and transpose operation, respectively. The off diagonal element of $\boldsymbol{\Sigma}$ represents the degree of correlation among the photons. Expression ``$(\mathbf{t} - \boldsymbol{\mu} - t_{\text{annih}})^T \mathbf{\Sigma}^{-1} (\mathbf{t} - \boldsymbol{\mu} - t_{\text{annih}})$'' represents the square of Mahalanobis distance between the observed ``$\boldsymbol{t}$'' and the mean ``$\boldsymbol{\mu}+t_{annih}$'' time vectors. Larger distance provide lower value of probability density which translates to inferior timing response for the pair detection elements. Moreover, if the observed photons are uncorrelated, the off diagonal elements of matrix $\boldsymbol{\Sigma}$ become zero, in this case equation (\ref{eq:probdensity}) reduces to the convolution of individual Gaussian functions, given as:
\begin{equation}
 P[\mathbf{t} \mid t_{\text{annih}}] = \prod_{j=1}^{n} \frac{1}{\sqrt{2\pi \sigma_{j}^2}} exp \Big( -\frac{(t_{j}-\mu_{j}-t_{annih})^2}{2\sigma_{j}^2} \Big )  
 \label{eq:gauss_mult}
\end{equation}
From the observed photon probability distribution (\ref{eq:probdensity}), the Fisher equation and the width of the observed light distribution can be calculated as :
\begin{align}
\begin{split}
    I(t_{\text{annih}}) = - E \left[ \frac{\partial^2}{\partial t_{\text{annih}}^2} \ln P(\boldsymbol{t} \mid t_{\text{annih}}) \right] 
= \mathbf{1}^T \boldsymbol{{\Sigma}^{-1}} \mathbf{1} = \sum_{i=1}^{n} \sum_{j=1}^{n} (\boldsymbol{{\Sigma}^{-1})_{ij}} \\
\end{split}\\
\begin{split}
\sigma_{lb}^2 = \frac{1}{I(t_{\text{annih}})}
\label{eq:cramer}
\end{split}
\end{align}
Here, $E[\cdot]$ denotes the expectation value of the argument, and $\mathbf{1}$ represents a vector of length $n$ with unity elements. Equation (\ref{eq:cramer}) represents the Cram\'er--Rao lower bound for timing resolution. Using Equation (\ref{eq:gauss_mult}), one can observe that for uncorrelated photons, the Fisher information matrix $I(t_{\text{annih}})$ becomes diagonal, with elements $1/\sigma_{j}^2$  along the diagonal and zeros elsewhere. This implies that greater Fisher information corresponds to narrower light collection profiles, thereby improving timing performance. The Cram\'er--Rao bound represents the fundamental limit on the precision with which a scintillator detector can estimate $t_{\text{annih}}$. The lower-bound standard deviation, $\sigma_{lb}$, depends on several factors namely: scintillation emission/decay characteristics, optical properties of light transport, crystal geometry along with surface reflectivity, and the timing dispersion of photosensor. When additional broadening effects—such as pulse height resolution and time jitter from the signal processing chain—are included, the overall timing resolution can be expressed as:

\begin{eqnarray}
    \sigma_{t}^2 = \sigma_{lb}^2 + \sigma_{resol.}^2
    \label{eq:totalbroad}
\end{eqnarray}
Value of $\sigma_{t}$ is estimated for a pair of detectors capturing two 511 keV gamma-rays to further evaluate the coincidence resolving time (CRT), given as:
\begin{equation}
    CRT (FWHM)= 2.35 \times\sqrt{\sigma_{t_{1}}^2 + \sigma_{t_{2}}^2+ \sigma_{jitter}^2 } =\frac{2 \;\Delta r}{c}
       \label{eq:crt}
\end{equation}
Where ``$c$'' is the speed of light, this equation is important in PET imaging since it is directly connected to the image resolution . In the subsequent sections, CRT value determined for LYSO and plastic scintillator detectors are discussed.
\section{Geant4 implementation of crystal bar and SiPM pad}
\label{sec:geant4_implement}
\begin{figure}[!b]
    \centering
    \includegraphics[scale=0.25]{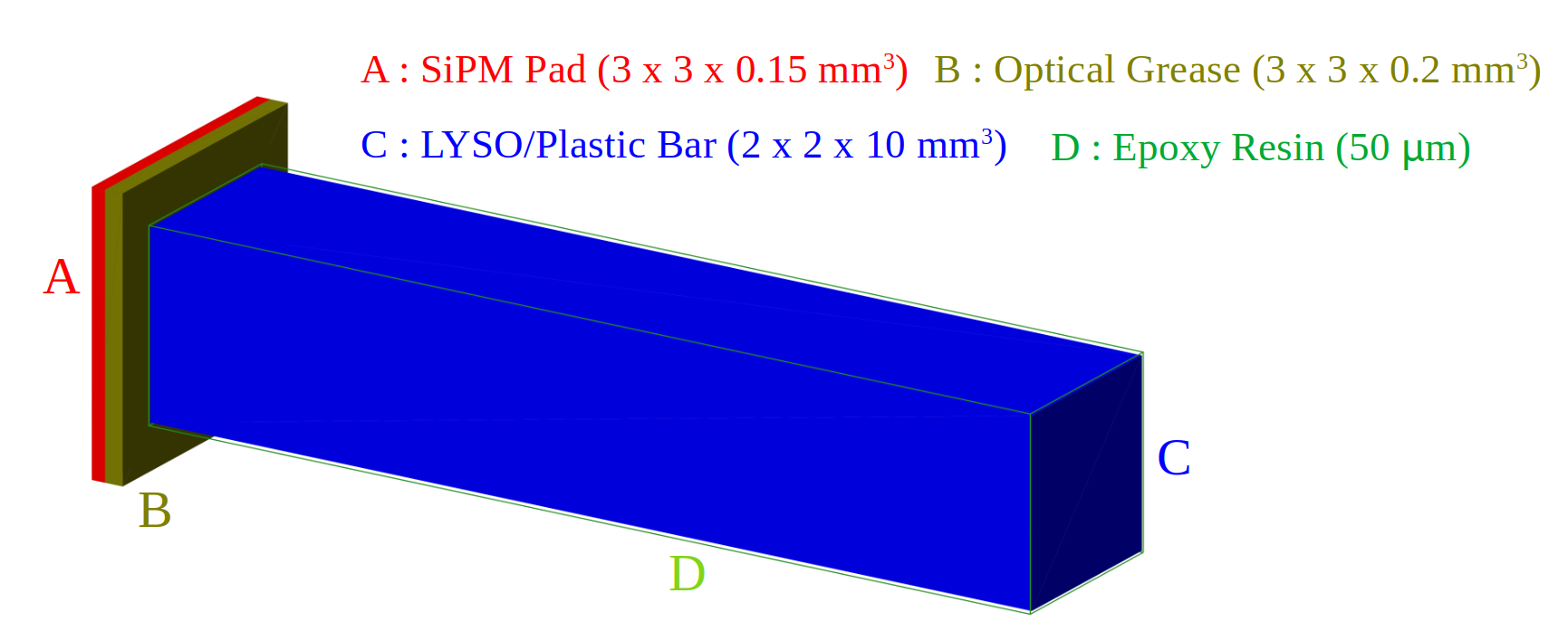}
    \caption{Assembly of scintillation bar with SiPM pad, constructed in Geant4. It is coupled to pad with a optical grease for a smooth light transmission. All the other surface is wrapped with a reflective epoxy resin layer of thickness 50 $\mu$m, shown as a wireframe. The surface roughness is also taken into account with optical properties mentioned in Table \ref{optical_table}. For the clarity of reader, different geometrical dimensions are mentioned in the legend.}
    \label{fig:barwithsipm}
\end{figure}

A pair of scintillator bars were modeled in Geant4 \cite{geant4} having version \texttt{4.10.07.p02}, positioned at a radial distance of 78 mm—a typical gantry radius for an animal PET scanner \cite{zaidiPET}. They were wrapped in a reflective coating of epoxy resin of thickness 50 $\mu$m. Crystal surface roughness was also taken into account by the micro facet angular dispersion parameter, $\sigma_{\alpha}$, along with different type of light reflection probability in Geant4. At the exit plane, photons first traversed a 200 $\mu$m optical grease layer before reaching a 150 $\mu$m thick SiPM wafer \cite{hamam}, as illustrated in Fig. \ref{fig:barwithsipm}. A 511 keV $\gamma$-ray point source was placed at the center between the two bars, emitting two back-to-back annihilated photons to mimic the radioisotope decay used in animal PET imaging. As soon as the interaction occurs inside the crystal volume, these $\gamma$-rays undergo Compton scattering or photoelectric processes, producing recoil electrons that generate scintillation flashes with characteristic temporal and energy distribution of the material. This light carried spatio-temporal information about the gamma-ray interaction point. The optical properties of the scintillator material were assigned via the G4Material property pointer in Geant4, with key parameters for LYSO and plastic bar materials is listed in Table \ref{tab:detprop}. To account for the surface roughness as well as the light reflection due to crystal/epoxy layer for the scintillation photons, different optical parameter were considered, as suggested by the literature \cite{Daneil2010}. They are listed in the table \ref{optical_table}. Light propagation through the crystal interface is modeled by surface finish of \texttt{groundbackpainted}  with a reflection probability of 94 \% offered by the epoxy resin. Different type of reflection probability like ; \texttt{specularlobe}, \texttt{specularspike}, and \texttt{backscattering} process were accounted,  mentioned as P(SL), P(SS), and P(BS) respectively in Table  \ref{optical_table}. Thus, photons undergo multiple reflections from their point of emission until their cumulative path length matches the material’s absorption length, in that case, they get absorbed, or they escaped through the optical grease. Using the \verb "G4UserStepping" class, the trajectory of each photon was tracked until the post-step volume was identified as the SiPM wafer with a quantum efficiency of 35 \% \cite{RVinke2014}.

To significantly reduce simulation time dominated by optical photon tracking, we leveraged Geant4's multithreading capability on an Intel 6-core (12th generation Core i5) desktop computer with 16 GB RAM running \texttt{Ubuntu 22.04.5 LTS}. The simulation employed \verb "G4MTRunManager", Geant4's specialized multithreaded run manager derived from \verb "G4RunManager", to enable parallel event processing distributed among 24 threads. In this configuration, while particle definitions and physics interactions were established globally for consistency, the master thread dynamically divided the event queue into manageable chunks that worker threads processed asynchronously. Each worker thread maintained independent instances of key user action classes - \verb "G4UserRunAction", \verb "G4UserEventAction", \verb "G4UserTrackingAction", and \verb "G4UserSteppingAction" - ensuring thread-safe operation. For optical photon analysis, tracking IDs recorded by the SiPM-sensitive detector are correlated with those in stepping actions to precisely determine the optical time spreads. All simulation data, including $\gamma$-ray interactions and light propagation details are systematically stored in ntuples using \verb "G4AnalysisManager". Upon completing all the events, the master thread coordinated with worker threads for the aggregation of statistics to generate final analysis output in ROOT \cite{Root} format, combining the distributed results into a unified data set.

%the aggregation of statistics from worker threads and generated final analysis output in ROOT format, combining the distributed results into a unified dataset.
\begin{table}[htbp]
    \centering
    \caption{Material properties of different scintillators used in Geant4 simulation}.
    \vspace{0.5cm}
    \label{tab:detprop}
    \begin{tabular}{l l l l}  % "l" for left-aligned columns; use "c" or "r" for center/right alignment
        \hline \hline 
        Parameter Name & LYSO \cite{RVinke2014, luxium, Zhang2008} &Plastic-Ooty \cite{ootyscint}&Plastic-BC404 \cite{knoll,luxium}\\ \hline \hline
        Density (gm/cm$^{3})$ & 7.1 & 1.032 &1.032  \\
        Refractive Index & 1.81 & 1.58 & 1.58 \\
        Absorption Length (cm)  & 1.2 &100 & 160 \\ 
        Light Yield (MeV$^{-1}$) & 27000 & 10000 &11832 \\
        Decay Time (ns) & 43 & 1.5 &1.5 \\
        $\lambda_{max}$ (nm) & 410 & 420 &405 \\
        H:C ratio & - & 1.1 &1.107\\
        \hline
        \hline
    \end{tabular}
    \label{tab:compare}
\end{table}

\begin{table}[htbp]
    \centering
    \caption{Optical properties of all the scintillators used in Geant4 simulation \cite{Daneil2010}}.
    \vspace{0.5cm}
    \label{tab:detprop}
    \begin{tabular}{l l}
    \hline
    \hline
     %\textcolor{blue}{Optical parameters}&\\
     Parameter Name  &Property/Value\\
     \hline
     \hline
      Surface model  &unified\\
      Surface finish& groundbackpainted \\
      Surface type & dielectric\_dielectric\\
      Reflectivity & 94\%\\
      P(SL) & 1\\
      P(SS) & 0\\
      P(BS) & 0\\
      $\sigma_{\alpha}$&4$^\circ$\\
      $n_{Grease}$ & 1.5\\
      $n_{Epoxy}$ & 1.53\\
       \hline
       \hline
    \end{tabular}
    \label{optical_table}
\end{table}
\begin{figure}[!h] %tbp]
 \centering
  \begin{tabular}{cc}
    \includegraphics[width=0.55\textwidth]{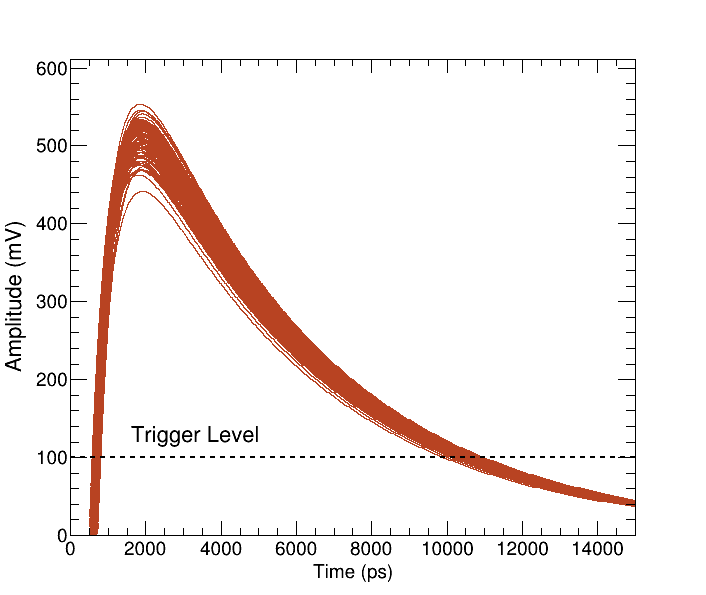} &
    \includegraphics[width=0.55\textwidth]{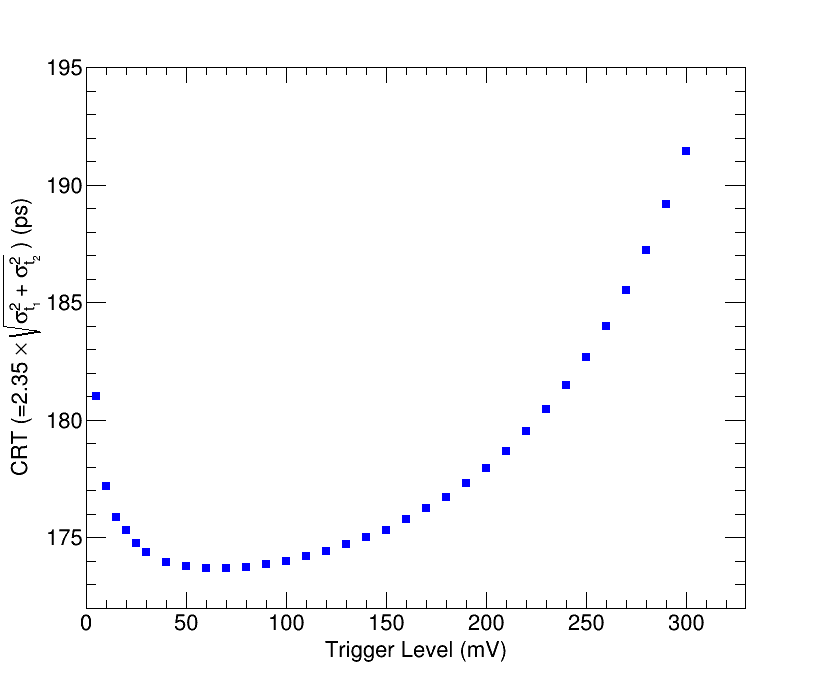} \\
  \end{tabular}
    \label{fig:tabular}
    \caption{Left : Collection of SiPM signal output, each is generated from the fastest 300 light photons reaches to the SiPM pad. Timestamp of each signal is calculated as soon as it crosses the trigger level. Right : Evolution of coincidence resolving time (CRT) as a function of trigger threshold  of SiPM signal.}
    \label{fig:sipmpulses}
\end{figure}

\section{SiPM signal modeling}
\label{sec:sipm_sig}
SiPM signal is calculated under the photopeak events for the first 300 light photons \cite{RVinke2014}. These photons gets collected at the SiPM wafer undergo multiple reflections within the crystal, resulting in an effective path length that contributes to the intrinsic time dispersion. From the emission point to the collection end, the total arrival time is given by:
\begin{eqnarray}
t_{arriv}   = t_{\gamma} + \Delta t_{Lt} + t_{ps}
\label{eqn:arrival}
\end{eqnarray}
Here, $t_{\gamma}$ represents the interaction time of the $\gamma$-ray photon in the crystal, which produces a recoil electron, obtained by using Geant4’s \verb "GetGlobalTime()" function. The term $\Delta t_{Lt}$ accounts for the light transit time , encompassing the duration from scintillation to the collection point. This includes effects such as crystal surface reflections, photon statistics, decay time variations, light transport correlations, self absorption, re-emission, and optical coupling efficiency with SiPM pad. The SiPM signal propagation time, $t_{ps}$, can be modeled by using a Gaussian probability density function:
\begin{eqnarray}
    PDF_{t_{ps}} = \frac{1}{\sqrt{2 \pi} \sigma_{ps}} \; exp \Big( -\frac{(t-\mu_{ps})^2}{2\sigma_{ps}^2} \Big)
    \label{sipmpdf}
\end{eqnarray}
where $\mu_{ps}$ is the mean arrival time and $\sigma_{ps}$ quantifies the time spread per photon. Since $\mu_{ps}$ does not contribute to pulse broadening, it is treated as a constant for simplicity. In the present calculation, single photon SiPM pad time spread, $\sigma_{ps}$, of 80 $ps$ is considered \cite{RVinke2014}. The value of $t_{ps}$ is sampled from $PDF_{t_{ps}}$ for each photon detected by the SiPM. Through optoelectronic conversion, each photon generates a signal at the collection end. In the present analysis, effects such as : dark current, optical crosstalk, afterglow, and Geiger-mode operation of the SiPM are not considered. The single photoelectron response (SER) function of SiPM pad can be modeled as a bi-exponential pulse \cite{RVinke2014}:

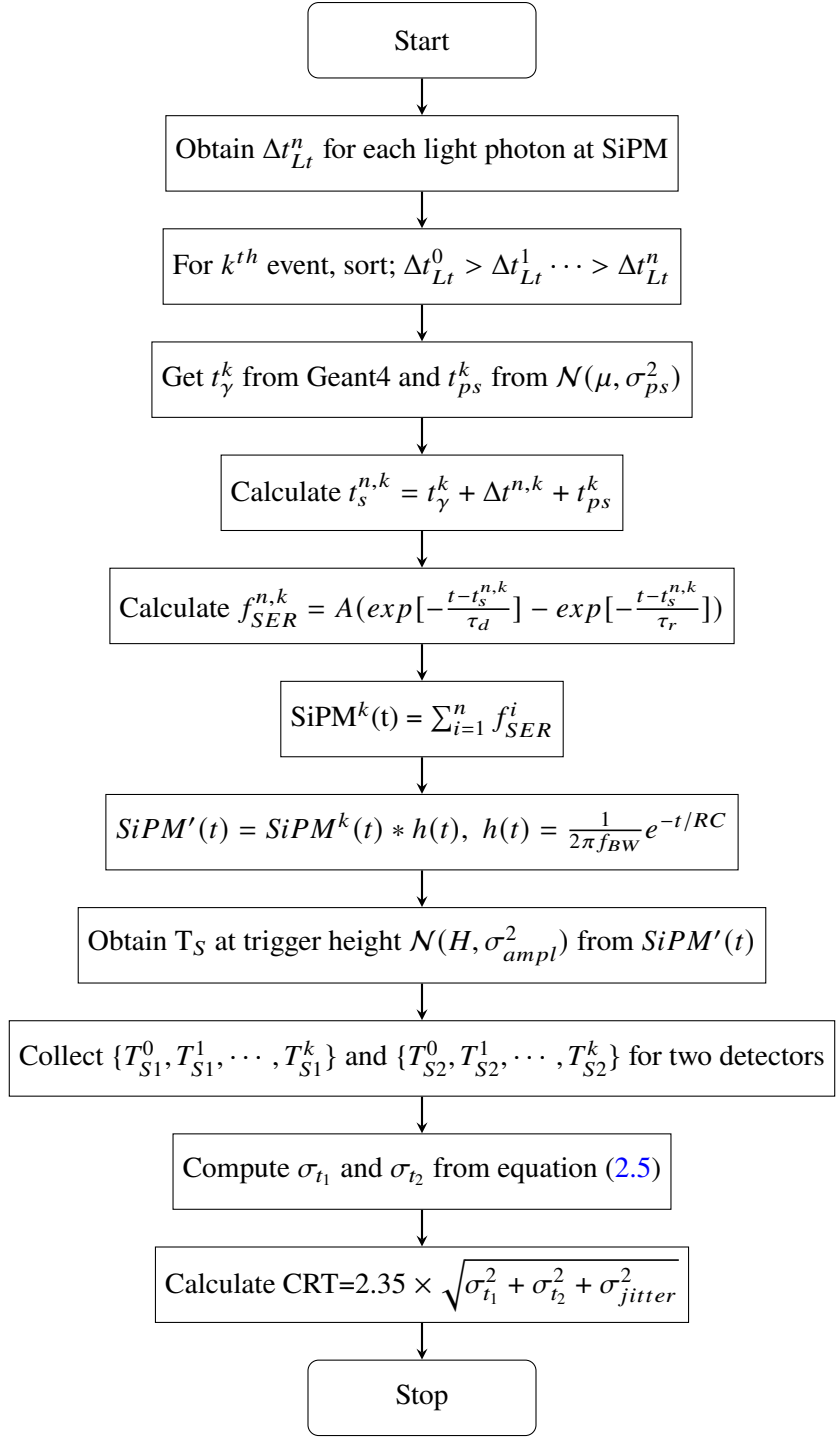
\begin{figure}[!t]
\centering
    \begin{tikzpicture}[
    node distance=1.5cm,
    startstop/.style={rectangle, rounded corners, minimum width=3cm, minimum height=1cm, text centered, draw=black, fill=white!30},
    process/.style={rectangle, minimum width=3cm, minimum height=1cm, text centered, draw=black, fill=white!20},
    decision/.style={diamond, minimum width=3cm, minimum height=1cm, text centered, draw=black, fill=green!30},
    io/.style={trapezium, trapezium left angle=70, trapezium right angle=110, minimum width=3cm, minimum height=1cm, text centered, draw=black, fill=yellow!30},
    arrow/.style={thick,->,>=stealth}
]

% Nodes
\node (start) [startstop] {Start};
\node (obtain) [process, below of=start] {Obtain $\Delta t_{Lt}^n$ for each light photon at SiPM};
\node (sort) [process, below of=obtain] {For $k^{th}$ event, sort; $\Delta t_{Lt}^0 > \Delta t_{Lt}^1 \cdots > \Delta t_{Lt}^n$};
\node (sort1)[process, below of=sort]{Get $t_{\gamma}^{k}$ from Geant4 and $t_{ps}^{k}$ from $\mathcal{N} (\mu, \sigma_{ps}^2)$};
\node (pulse) [process, below of=sort1] {Calculate $t_s^{n,k} = t_{\gamma}^{k} + \Delta t^{n,k} + t_{ps}^{k}$};
\node (pulse1) [process, below of=pulse] {Calculate $f_{SER}^{n,k}=A(exp[-\frac{t-t_{s}^{n,k}}{\tau_{d}}] -exp[-\frac{t-t_s^{n,k}}{\tau_r}])$};    
\node (signal) [process, below of=pulse1] {SiPM$^{k}$(t) = $\sum_{i=1}^{n} f_{SER}^i$};
\node (convolve)[process, below of=signal]{$SiPM'(t)=SiPM^{k}(t) * h(t), \; h(t) = \frac{1}{2\pi f_{BW}} e^{-t/RC}$};
\node (out) [process, below of=convolve] {Obtain T$_S$ at trigger height $\mathcal{N} (H, \sigma_{ampl}^2)$ from $SiPM'(t)$};
\node (time) [process, below of=out] {Collect \{$T_{S1}^{0}, T_{S1}^{1}, \cdots ,T_{S1}^{k}\}$ and \{$T_{S2}^{0}, T_{S2}^{1}, \cdots, T_{S2}^{k} \}$ for two detectors};
\node (sigma1) [process, below of=time] {Compute $\sigma_{t_{1}}$ and $\sigma_{t_{2}}$ from equation (\ref{eq:totalbroad})};% \{$T_{S}^{j,1}$\} and \{$T_{S}^{j,2}$\}};
\node (sigma2) [process, below of=sigma1] { Calculate CRT=2.35  $\times \; \sqrt{\sigma_{t_{1}}^2 + \sigma_{t_{2}}^2 + \sigma_{jitter}^2}$ };
\node (stop) [startstop, below of=sigma2] {Stop};

\draw [arrow] (start) -- (obtain);
\draw [arrow] (obtain) -- (sort);
\draw [arrow] (sort) -- (sort1);
\draw [arrow]  (sort1) -- (pulse);
\draw [arrow] (pulse) -- (pulse1);
\draw [arrow] (pulse1) -- (signal);
\draw [arrow] (signal) -- (convolve);
\draw [arrow] (convolve) -- (out);
\draw [arrow] (out) -- (time);
\draw [arrow] (time) -- (sigma1);
\draw [arrow] (sigma1) -- (sigma2);
\draw [arrow] (sigma2) -- (stop);

\end{tikzpicture}
\caption{Flowchart of SiPM signal calculation used to determine the CRT value of two bars.}
\label{fig:flowchart}
\end{figure}

\begin{figure}[!b]
\small
 \centering
  \begin{tabular}{cc}
    \includegraphics[width=0.55\textwidth]{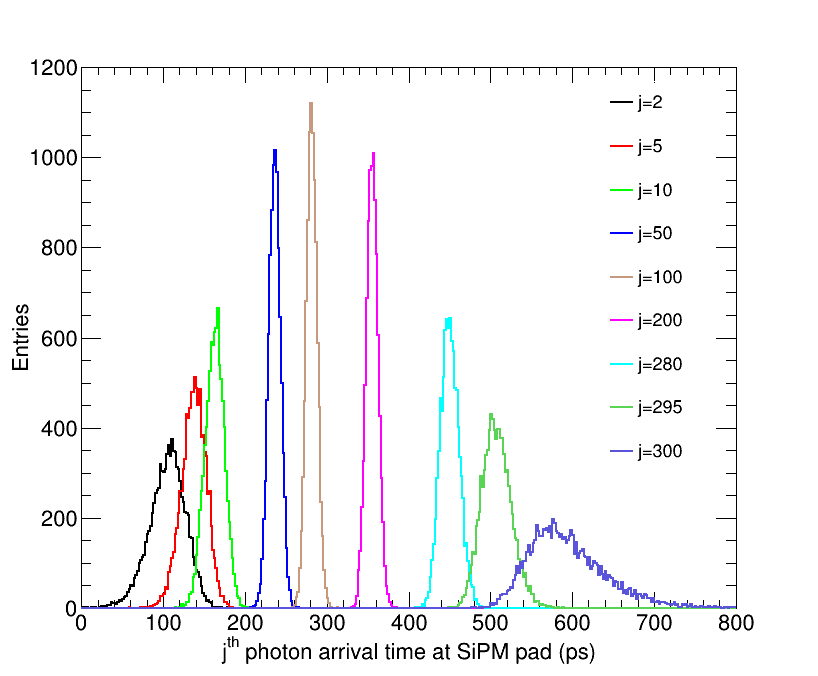} &
    \includegraphics[width=0.55\textwidth]{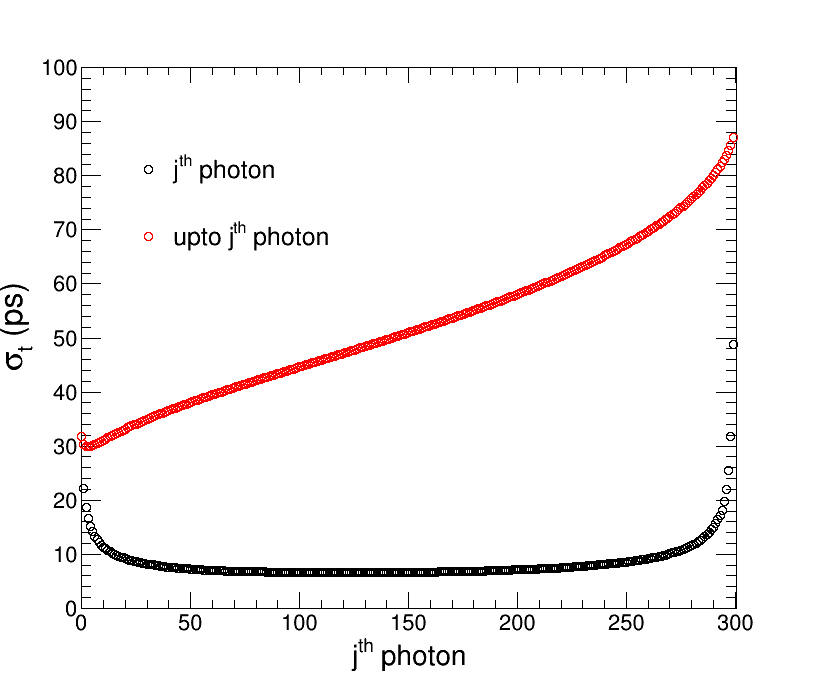} \\
  \end{tabular}
    \caption{Left : Arrival time distribution of various light photons under photopeak events of a LYSO bar. Right : The graph shows the variation in the width of the j$^{th}$ photon arrival time and the accumulated width, calculated for  a full energy $\gamma$-ray event.}
    \label{fig:photon_time}
\end{figure}

\begin{eqnarray}
    f_{SER} = A. \Big [ exp \Big( -\frac{(t-t_{arriv})}{\tau_{d}} \Big) - exp \Big (-\frac{(t-t_{arriv})}{\tau_{r}} \Big)  \Big]
\end{eqnarray}
In the present simulation, the rise time ($\tau_{r}$) is set to be 500 ps, while the decay time ($\tau_{d}$) is taken as 5 ns \cite{RVinke2014}.  First, the detected photons arrival times are sorted in ascending order followed by SiPM timing dispersion using Equation (\ref{eqn:arrival}). A bi-exponential pulse, $f_{SER}(t)$, is then generated for each photon arrival (servived  after 35 \% quantum  efficiency of SiPM). For each event, the fastest 300 photons pulse responses are then summed to produce the SiPM signal. For simplicity, we assigned a value of 511 mV height to the photopeak, as the typical signals from the SiPM front-end electronics are several hundred millivolts high. To include the  detector pulse height broadening effect on time dispersion at photopeak energy of 511 keV $\gamma$-rays, an energy resolution of 11 \% was introduced \cite{RVinke2014} in the pulse height. We also have considered the effect of the system's 1 GHz bandwidth, imposed by either the front-end electronics or the oscilloscope, on signal acquisition. The expected signal was obtained by convolving the input with a low-pass filter kernel. The resulting waveforms are illustrated in Fig. \ref{fig:sipmpulses} with a peaking time of 1.5 ns. In order to determine the timestamp of the detected events, a time pick-off method is applied to each generated signal. A trigger level is therefore adjusted at a specific height which is determined by the Gaussian sampling of 5 mV noise ($\sigma$) introduced in the trigger level  above the pulse baseline to record the pick-off time whenever the signal crosses this level. This process is repeated for a pair of LYSO bars to collect the sets of timestamps ; $\{ T_{s1}^{k}\}$ and $\{ T_{s2}^{k}\}$, for large number of events (k). Signals are generated for 4,615 full-energy, coincident $\gamma$-ray events. For each event, at least 300 optical photons survived after accounting for the quantum efficiency and reached the SiPM wafer in Geant4. Standard deviations ($\sigma_{t_{1}}$ and $\sigma_{t_{2}}$) are then computed to derive the CRT values. They are further propagated with signal readout electronics timing dispersion of 57 ps (FWHM) \cite{Fallu2007} to mimic the experimental resolving time. Obtained locus of CRT at different trigger level is depicted in Fig. \ref{fig:sipmpulses} (right). It shows the minima of curvature at $\approx$ 174 ps which is close to 175 ps, reported earlier for the same dimension \cite{Auffray2011}. Higher values of CRT at lower trigger level indicates that the noise contribution in the signal processing chain worsens the timing resolution. Whereas, the nonlinear increase in the CRT values at higher trigger levels suggests larger sensitivity towards the signal height. For the better clarity of the procedure, a flowchart of the methodology is provided in Fig. \ref{fig:flowchart}. The statistical distribution of optical photons hit times at the SiPM pad is shown in Fig. \ref{fig:photon_time} (left), representing the collection for the $j^{th}$ photons.   The arrival time of each photon exhibits a spread as a result of the convolution of the crystal’s timing response and the single electron response (SER) broadening of the pad. Notably, photons with lower and higher indices experience greater dispersion compared to those in the central region, a trend consistent with prior findings \cite{RVinke2014}. The arrival time dispersion, including SiPM response, ranges from 22 ps to 50 ps, with a central plateau of about 7 ps, as illustrated in Fig. \ref{fig:photon_time} (right). This behaviour provides key insights into the scintillation characteristics of LYSO. Photons with lower $j$ indices correspond to the rise time of LYSO’s temporal response and are primarily influenced by the stochastic nature of 511 keV interactions and fluctuations in photon yield. In contrast, mid-index photons originate predominantly from the stable decay region of the scintillation process, resulting in minimal dispersion. Meanwhile, high-index photons exhibit poorer timing resolution due to decay time fluctuations and are further broadened by optical reflections. The cumulative time broadening across all 300 photons is also computed, revealing a maximum broadening of \textcolor{blue}{88 ps}, as depicted in Fig. \ref{fig:photon_time} (right).

\section{Analysis with plastic scintillator bars}
\label{sec:plastic_analysis}
\begin{figure}[!t]
    \centering
    \includegraphics[width=0.55\linewidth]{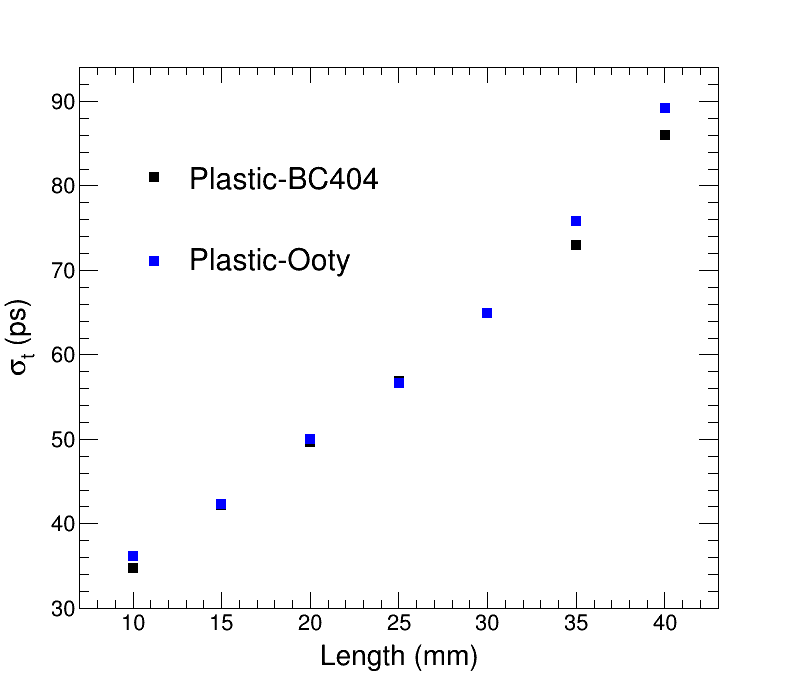}
    \caption{Contribution to the time broadening by the fastest collected photon at different bar lengths.}
    \label{fig:sigmat}
\end{figure}

The Geant4 simulation framework was used to analyze plastic scintillator bars (BC-404 and Ooty) by substituting their material properties in place of LYSO crystals. Given that plastic scintillators consist of hydrocarbon compounds, their lower effective atomic number (Z) and density compared to LYSO result in significantly fewer interactions with 511 keV gamma rays. Owing to their reduced photoelectric cross-section, no photopeak events are detected—only a Compton continuum is observed. The study employed pairs of plastic bars (2 mm $\times$ 2 mm cross-section, variable lengths) irradiated with back-to-back 511 keV gamma rays from a point source. Each Compton interaction produces a recoil electron, with its energy converted into scintillation photons as per the light yields listed in Table \ref{tab:compare}. Both plastic scintillators exhibits faster decay times (1.5 ns) than LYSO (43 ns), therefore relatively faster SiPM pulses can be expected . However, their higher optical absorption lengths lead to relatively increased optical reflections. Under identical optical surface properties as LYSO, the simulation tracked photons escaping the bars and reaching the SiPM pad. Due to the plastic's longer attenuation lengths, the collected photons displays broader optical time dispersion compared to LYSO. It has been observed that for energy threshold of 25 keV, maximum number of photons received at SiPM pad under Compton edge condition is $\sim$ 1000, whereas the miniumum number of photons for a bar length 45 mm (10 mm) is about 50 (250). Suggesting, that the optical photon yeild at the SiPM wafer depends strongly on the $\gamma$-ray interaction depth inside the crystal volume. Thus, more optical loss is observed for the peripheral interactions compared to those which are closer to SiPM pad. In every instance of gamma-ray interaction, the first photon to reach the SiPM pad—corresponding to the shortest effective tracking path or earliest arrival time—triggers the signal \cite{Moskal2016}. Figure \ref{fig:sigmat} illustrates the optical time spread observed by the SiPM pad for the fastest photon across different scintillator bar lengths. This spread was determined by analysing the arrival times of the first upto 300 photons $\Delta t_{Lt}$ from Geant4 simulations. The photons were sorted chronologically, and the earliest arrival time was identified. This value was then added with the SiPM timing dispersion sampled from equation (\ref{sipmpdf}). The resulting time spread, plotted as a function of the bar length in Figure \ref{fig:sigmat}, reveals a near-linear relationship for both types of plastic scintillators. These values represent the fundamental lower limit of optical time spread achievable for a given combination of scintillator material and photosensor characteristics.

\begin{figure}[!t]
  \centering
  \begin{tabular}{cc}
    \includegraphics[width=0.55\textwidth]{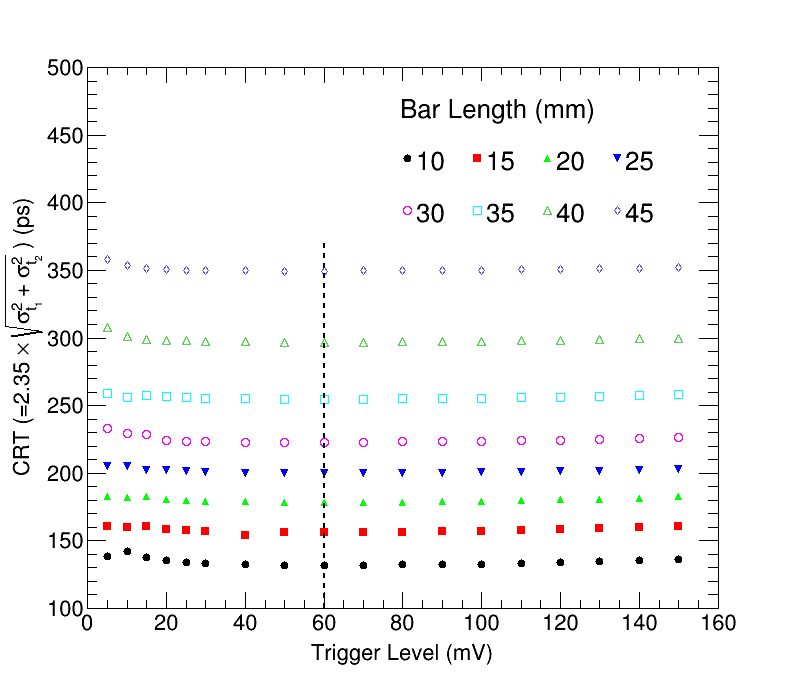} &
    \includegraphics[width=0.55\textwidth]{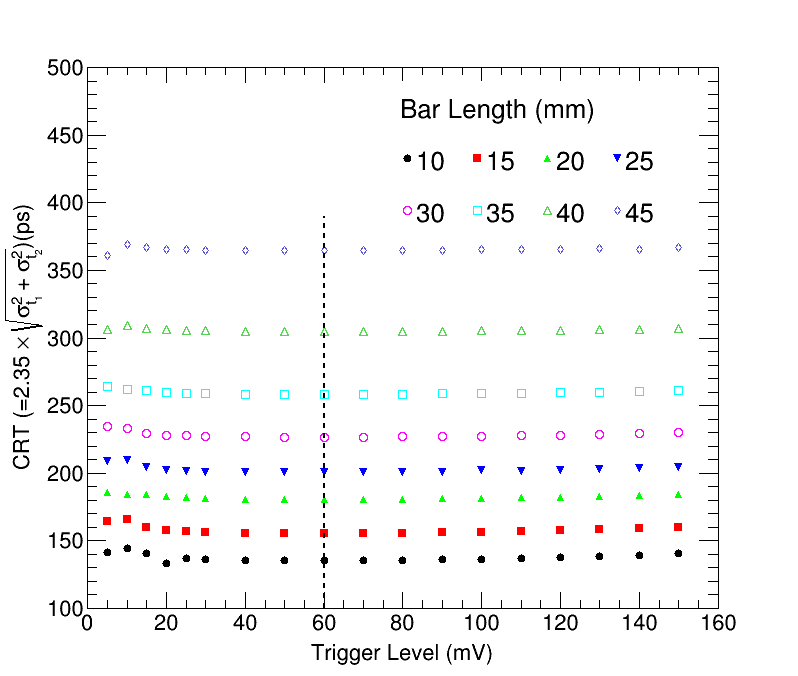} \\
  \end{tabular}
  \caption{Left : CRT values obtained as a function of different trigger level for different length of plastic-BC404 scintillator bar is plotted. Right : same as left graph plotted for Plastic-Ooty scintillator. }
  \label{fig:crt_bothplastics}
\end{figure}

\begin{figure}[!t]
  \centering
  \begin{tabular}{cc}
  \includegraphics[width=0.55\textwidth]{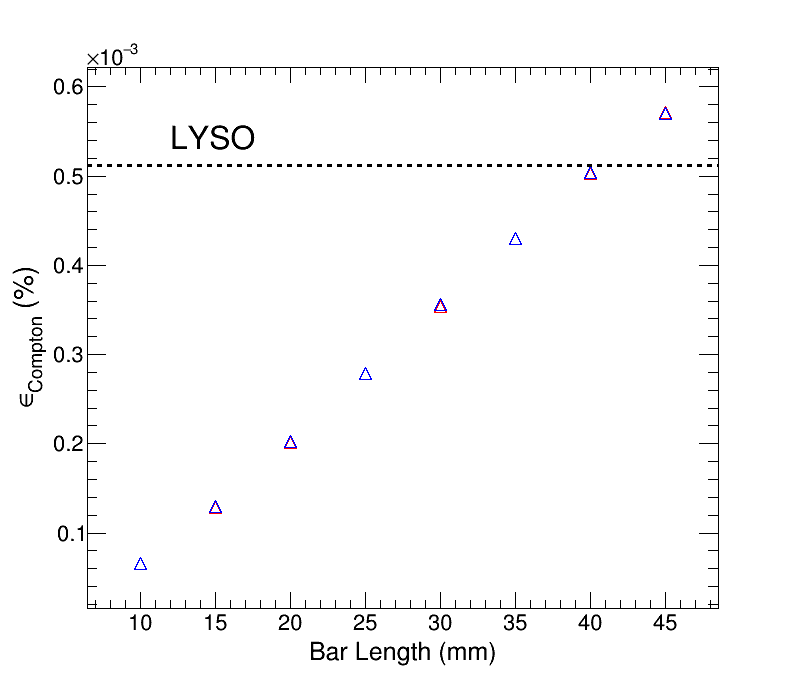}&
    \includegraphics[width=0.55\textwidth]{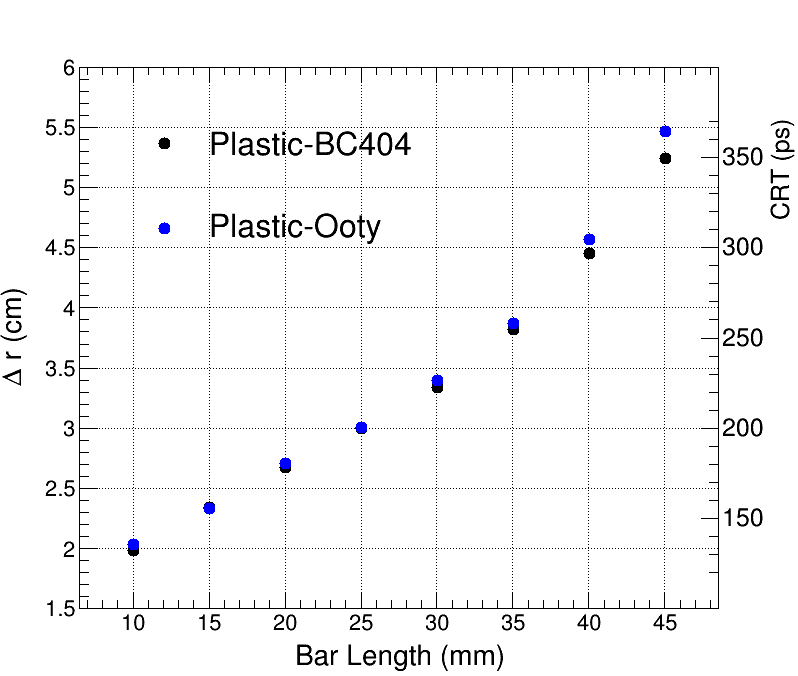} \\
  \end{tabular}
  \caption{Left : Detection efficiency for BC404 (blue triangle) and Ooty (red traingle) is plotted for different bar length. For comparison, the photopeak detection efficiency of LYSO bar having dimensions of 2 mm $\times$ 2 mm $\times$ 10 mm  is shown by dotted line. Right : CRT value translation to spatial imaging resolution for different length of bar is depicted for both the scintillators.}
  \label{fig:eff_deltar_graph}
\end{figure}
The CRT values were evaluated across various scintillator bar lengths by generating SiPM signals for the fastest detected photons using Equation (\ref{eq:crt}). Height of signal (in mV) is generated that corresponds to the recoil energy of electron. Event-by-event analysis was conducted for all the Compton scattering events with recoil electron energy exceeding 25 keV. To incorporate detector response variations, the energy resolution corresponding to maximum recoil electron energy deposition \cite{Moskal2015} is applied to the SiPM bi-exponential pulses. Their timestamps were determined at various trigger threshold values. The ensembles of timestamps were collected for both the scintillator bars to calculate the CRT values as obtain from the algorithm mentioned in Fig. \ref{fig:flowchart}. For different trigger level, the trend of CRT for both type of plastic scintillators are portrayed in Fig. \ref{fig:crt_bothplastics}.  Results demonstrate a nonlinear CRT locus, reflecting sensitivity to signal pulse shape, and consistent with prior observations \cite{RVinke2014}. It is evident from the graphs that beyond the length of 35 mm, the curves separation relatively increases, attributed to larger time broadening carried by the light photon at the exit plane. The locus minima is found to be at the trigger level 60 mV. We further evaluates the required plastic scintillator bar length needed to match both the imaging resolution and detection efficiency of LYSO crystals. Simulations determined the integrated Compton detection efficiency using a 25 keV energy threshold, calculated by the following formula:
\begin{equation}
    \epsilon_{Compton} = \frac{\mbox{Number of events in Compton Continuum} }{\mbox{Number of events emitted Isotropically}}
\end{equation}
Figure \ref{fig:eff_deltar_graph} presents these efficiency values across varying bar lengths, showing a linear increase with bar volume due to enhanced gamma-ray interaction probability. The results demonstrate that plastic scintillators require approximately four times the length of standard LYSO crystals to achieve comparable detection efficiency of $\approx$ 0.5 $\times \; 10^{-3}$ \%, as indicated by the horizontal dotted line representing LYSO's photopeak efficiency. However, this gain in efficiency comes with a trade-off in timing performance $\approx$ 300 ps for plastics versus 174 ps for LYSO—attributed to the combined effects of optical path length and characteristics temporal response in plastic materials. Image resolution for different bar length is depicted in Fig. \ref{fig:eff_deltar_graph} (right). It can be inferred that for equivalent imaging resolution $\Delta r$ (derived from $\Delta r=\frac{c \Delta t}{2}$), a 20 mm plastic bar length suffices, though this configuration yields 40 \% of  the integrated Compton efficiency compared to LYSO photopeak efficiency.

\section{Discussions and future outlook}
\label{sec:discussion}
Presented Geant4 simulations evaluate plastic scintillators as potential alternatives to LYSO crystals, determining the required equivalent plastic bar lengths. While traditional PET scanners never utilised plastic scintillators due to their lower gamma-ray interaction probability (leading to longer scan times), these materials offer compelling advantages: superior volume scalability, cost per unit volume is 80 times less than Lu-based crystals \cite{Moskal2012}, approximately 100 times longer attenuation lengths, and an order of magnitude faster decay times compared to LYSO. While the photon count incident on the SiPM is limited to 1000 photons, a timing resolution equivalent to LYSO is attainable with a 20 mm bar length. To cope up with low detection efficiency, the bar length must be increased to 40 mm,   which introduces  a natural trade-off by degrading the timing resolution to 300 ps. Crucially, our analysis reveals comparable performance between TIFR Ooty's in-house developed plastic material and commercial BC404 scintillators. These findings open new possibilities to innovate animal PET gantry designs that will be applicable to both custom and commercial plastic scintillators. Future work will be focus on MOBY phantom \cite{moby} imaging studies by using 2 mm $\times$ 2 mm $\times$ 40 mm bars arranged in various mesh designs to form a animal PET module that will be further arranged in various gantry configurations to optimize the performance.
\section{Acknowledgment}
One of the authors (DS) acknowledges the Central University of Kashmir for providing a high performance desktop computer to perform the optical simulations presented in the article.

% Bibliography

%% [A] Recommended: using JHEP.bst file
%% \bibliographystyle{JHEP}
%% \bibliography{biblio.bib}

%% or
%% [B] Manual formatting (see below)
%% (i) We suggest to always provide author, title and journal data or doi:
%% in short all the informations that clearly identify a document.
%% (ii) please avoid comments such as "For a review'', "For some examples",
%% "and references therein" or move them in the text. In general, please leave only references in the bibliography and move all
%% accessory text in footnotes.
%% (iii) Also, please have only one work for each \bibitem.

\end{document}